\documentclass{PoS}
\usepackage{amsmath}
\usepackage[utf8x]{inputenc}

\newcommand{\eq}{\begin{equation}}
\newcommand{\eqx}{\end{equation}}
\newcommand{\eqn}{\begin{eqnarray}}
\newcommand{\eqnx}{\end{eqnarray}}
\newcommand{\alg}{\begin{align}}
\newcommand{\algx}{\end{align}}

\newcommand{\f}[2]{\frac{#1}{#2}}

\newcommand{\Dl}{\Delta}
\newcommand{\bt}{\beta}

\newcommand{\al}{\alpha}
\newcommand{\eps}{\varepsilon}
\newcommand{\nn}{{\cal N}}

\newcommand{\qqqq}{\quad\quad\quad\quad}

\newcommand{\OO}[1]{{\cal O}\left(#1\right)}
\newcommand{\cor}[1]{\left\langle#1\right\rangle}

\newcommand{\teff}{T_{eff}}

\title{Strongly coupled plasma - hydrodynamics, thermalization and nonequilibrium behavior}

\ShortTitle{Strongly coupled plasma - hydrodynamics, thermalization and nonequilibrium behavior}

\author{\speaker{Romuald JANIK}\\
        Jagiellonian University, Kraków\\
        E-mail: \email{romuald@th.if.uj.edu.pl}}

\abstract{In this talk I will describe various features of time-dependent dynamics of strongly coupled plasma from the perspective of the AdS/CFT correspondence. I will 
take as an example boost-invariant plasma flow and concentrate on the properties of hydrodynamic expansion, thermalization versus hydrodynamization and some features of nonequilibrium behaviour.}

\FullConference{KMI International Symposium 2013 on “Quest for the Origin of Particles and the Universe"\\
                 11-13 December, 2013\\
                 Nagoya University, Japan}

\begin{document}

\section{Introduction and motivation}

One of the outstanding theoretical problems, relevant for
the experimental heavy-ion programs at LHC and RHIC, is the understanding of real-time
dynamics of far-from-equilibrium quark-gluon plasma.
This physics is especially elusive, as the quintessential tool of nonperturbative
QCD --- lattice QCD cannot be applied, since it is formulated in Euclidean
signature. 

In this talk I will concentrate on an approach to these kind of problems
through the AdS/CFT correspondence \cite{adscft}, which postulates 
an equivalence between $\nn=4$ supersymetric gauge theory and superstring
theory in $AdS_5 \times S^5$ spacetime. The reasons which make this
correspondence an effective tool for studying the relevant physics are
twofold. It is formulated directly in Minkowski signature and moreover,
at strong coupling, the dual description essentially reduces to
classical gravity for which we have effective calculational tools.
Of course, the downside is that we are no longer dealing directly 
with QCD but rather with a different gauge theory.
Fortunately, once we are studying plasma physics (i.e. QCD physics
for temperatures above the confinement-deconfinement phase transition/crossover),
the chief differences from QCD go away. In particular, supersymmetry
is explicitly broken by the temperature, which also becomes the dominant
scale. Indeed, recent investigation of transport and kinetic equations
in $\nn=4$ SYM indicate that the differences with analogous
computations in QCD amount essentially just to a difference
in the number of degrees of freedom \cite{Czajka}.

However, having said all this, one has to nevertheless keep in mind
some differences which may be relevant for various concrete questions.
One key difference is the absence of a phase transition in $\nn=4$ SYM.
Consequently, the plasma expansion and cooling will continue indefinitely.
Secondly, the coupling in $\nn=4$ SYM is not running and has to be large
in order for the AdS/CFT methods to be effective in the plasma
context. So in particular mixed soft/hard physics may be significantly different.
Therefore, the (qualitative) relevance of particular computations in $\nn=4$ SYM
for QCD problems has to be judged depending on the concrete
physical question considered. Moreover, one should eventually consider
passing to more complicated AdS/CFT setups which involve gauge theories
which might be closer to QCD.

Before I finish this brief introduction, I would like to emphasize
that the AdS/CFT correspondence provides for us a concrete example
of a strongly coupled gauge theory in which we may study in very much
detail nonperturbative real-time dynamics.
We may then confront our theoretical expectations and prejudices with these results
and investigate possible relevance for realistic quark-gluon-plasma dynamics.
One may also hope to obtain certain quantitative answers
which may serve as a useful reference point for QCD applications 
-- the key example is the famous shear viscosity to entropy ratio \cite{shear1,shear2}. 

\section{Boost invariant flow}

In this talk I will review our work on exploring the interrelations between
hydrodynamic and nonequilibrium behaviour within the context of a boost-invariant
plasma configuration. Such a setup is a very appealing laboratory for exploring
nonequilibrium physics since the symmetry assumptions significantly reduce
the complexity of the problem. Then the dual AdS gravitational description
is essentially two-dimensional, which simplifies the numerical solution of Einstein's
equations. At the same time, the assumed symmetry is not overconstraining
the system and almost all of the physics relevant for thermalization
is still present. Indeed, we may and do observe complicated nonequilibrium
dynamics at early times and a gradual transition to hydrodynamic evolution.

These symmetry assumptions were first proposed by Bjorken in \cite{Bjorken}
as an approximate description of a plasma fluid produced in a a very
highly energetic collision. Thus, following Bjorken, we assume invariance
under boosts in the longitudinal plane. Moreover, we assume unifomity
in the transverse directions. With these assumptions in place,
the conservation of energy-momentum and the equation of state (which for 
the $\nn=4$ theory is just the tracelessness condition of $T_{\mu\nu}$)
implies that all components of $T_{\mu\nu}$ can be expressed
in terms of just a single function $\eps(\tau)$, which is the
energy density at mid-rapidity. In particular, the longitudinal and transverse pressures
(defined through the appropriate diagonal components of $T_{\mu\nu}$)
are given by
\eq
p_L=-\eps -\tau \f{d}{d\tau} \eps \qqqq
p_T=\eps + \f{1}{2}\tau \f{d}{d\tau} \eps\,
\eqx  
In the following, it turns to be convenient to use the notion of \emph{effective
temperature} $\teff$ which is defined to be the temperature of a thermal system
in $\nn=4$ SYM with energy density equal to the instantenous energy density 
$\eps(\tau)$:
\eq
\eps(\tau) = \f{3}{8} N_c^2 \pi^2 \teff^4(\tau)
\eqx
In this way one can get rid of the trivial differences between
gauge theories due to various numbers of degrees of freedom.

\section{Hydrodynamics in the boost invariant setting}
\label{s.hydro}

Let us now revisit hydrodynamics in the boost invariant setting.
Hydrodynamics amounts to the assumption that the whole long wavelength
dynamics is completely expressible in terms of a local flow velocity
and local temperature\footnote{Here I assume the conformal equation of state $E=3p$ 
($T^\mu_\mu=0$).}.
Then the energy-momentum tensor $T_{\mu\nu}$ is expressed in an expansion in the number of
derivatives of the flow velocity.
\eq
\label{e.hydro}
T^{\mu\nu}_{hydro}=\f{1}{3}\eps \left( 4 u^\mu u^\nu+ \eta^{\mu\nu} \right)+
\eta \left( P^{\mu \al} P^{\nu \bt} \partial_{(\al} u_{\bt)} -\f{1}{3} P^{\mu\nu}
\partial_\al u^\al \right) +\ldots
\eqx
The leading term corresponds to a perfect fluid, the $1^{st}$ correction
is just the shear viscosity multiplying the standard shear tensor
and subsequent terms would involve new, higher
order transport coefficients and new tensorial structures with more than one derivative
of the flow velocity $u^\mu$ (see~\cite{BAIER,FLUIDGRAVITY} for explicit expressions).
Hydrodynamic equations of motion are first order equations given
just by energy-momentum conservation\footnote{I am not discussing here 
the Israel-Stewart formulation.}
\eq
\partial_\nu T^{\mu\nu}_{hydro}=0
\eqx

The expansion (\ref{e.hydro}) can, in fact, be rigorously derived from
AdS/CFT \cite{FLUIDGRAVITY}. This procedure is usually called fluid/gravity
duality, although this name may be slightly misleading as gravity within AdS/CFT
contains a much richer range of dynamics than just (generalized) fluid
dynamics.

In the boost invariant setting (with no transverse dependence), the flow velocity
is completely fixed by symmetry and the only degree of freedom is the dependence
of the temperature on the (longitudinal) proper time $T(\tau)$. 
Let us note that this $T(\tau)$ would coincide with our definition
of effective temperature $\teff$ in the previous section.
The hydrodynamic
equations of motion become just a $1^{st}$ order (nonlinear) ODE for $T(\tau)$.

These simple facts have significant consequences. They imply that once
$T(\tau)$ has some fixed value at a given time, the rest of the evolution of $T(\tau)$
is completely fixed\footnote{In an Israel-Stewart formulation one may include one other
initial condition.} and unique. This conclusion holds irrespective of any
precise knowledge about the higher order hydrodynamic terms in the derivative
expansion (\ref{e.hydro}).

Alternatively, once we fix the late-time asymptotics of $T(\tau)$ through
\eq
\label{e.Tnorm}
\pi T(\tau) \sim \f{1}{\tau^{\f{1}{3}}}
\eqx
with a given fixed coefficient in the numerator,
the hydrodynamic evolution of a plasma system would be a \emph{unique} fixed
curve $T_{hydro}(\tau)$. Any deviations from this curve would be due to
the presence other \emph{non-hydrodynamic} excitations of the strongly coupled
plasma system. In particular, one could imagine preparing the plasma system
at $\tau=0$ in various ways, giving various initial conditions to these
non-hydrodynamic degrees of freedom\footnote{Of course, all these degrees of freedom
are expected to interact nonlinearly between themselves.}. Then we would
expect to observe a plethora of various curves $\teff(\tau)$ which would
converge to the single hydrodynamic curve $T_{hydro}(\tau)$.

Another interesting issue is what exactly is the shape of the 
hydrodynamic curve $T_{hydro}(\tau)$. The answer is of course well known
if we choose to restrict ourselves just to $1^{st}$ order viscous hydrodynamics
(or up to $3^{rd}$ order in the boost-invariant case). However, as we
go to earlier and earlier times, dissipation becomes more and more important
and higher order terms in the hydrodynamic derivative expansion in (\ref{e.hydro})
become relevant. Within the AdS/CFT correspondence, we recently obtained
numerically around 240 subsequent terms in the (boost-invariant) hydrodynamic
expansion \cite{HJW3}. It turns out that the hydrodynamic series 
has zero radius of convergence and is only asymptotic. However, there
are strong indications that it is Borel summable. Currently we are investigating
approximate Borel resummations which may serve as a benchmark of
resummed $T_{hydro}(\tau)$.

The time after which the relevant curve $\teff(\tau)$ would approach $T_{hydro}(\tau)$
would be the hydrodynamization time -- the time after which a purely hydrodynamic
description of the plasma would be valid. Knowing the energy-momentum tensor, allows us to ask whether the diagonal components
(i.e. longitudinal and transverse pressures) are approximately isotropic
which would be a necessary condition for local thermal equillibrium.

One of the aims of the AdS/CFT
investigations of the boost-invariant plasma system was to verify whether
the above physical picture indeed holds, and to understand
its fine details.

\section{The AdS/CFT description}

As the AdS/CFT correspondence is expected to be an equivalence between $\nn=4$ SYM theory
and superstring theory in $AdS_5 \times S^5$, each gauge theory state/phenomenon
should have a counterpart on the dual side. We thus seek to describe 
the time-dependent evolving strongly coupled plasma system in terms
of the dual degrees of freedom on the string side of the correspondence.
At strong (gauge theory) coupling, there appears a huge mass gap between
gravity modes in $AdS_5 \times S^5$ and degrees of freedom corresponding to
massive string states. Hence one may expect to describe the evolving plasma system
purely within gravity\footnote{Let us note that this does not hold at weak
coupling, where the massive string states are equally important as gravity.
Hence we lack a dual description of weakly coupled plasma}.
A proposal for doing this was formulated in~\cite{JP1}.
Some additional assumptions amounting to not turning on any exotic charges in $\nn=4$ SYM,
allow us to reduce the description to a 5D geometry:
\eq
ds^2=\f{g_{\mu\nu}(x^\rho,z)\, dx^\mu dx^\nu +dz^2}{z^2} \equiv g^{5D}_{\al\bt} dx^\al dx^\bt
\eqx
where $z\geq 0$ is the $5^{th}$ coordinate, with the $AdS$ boundary given by $z=0$.
Its time evolution is given by 5D \emph{vacuum} Einstein's equations with negative cosmological
constant:
\eq
R_{\al\bt}-\f{1}{2}g^{5D}_{\al\bt} R - 6\, g^{5D}_{\al\bt}=0
\eqx
Given the 5D geometry $g^{5D}_{\al\bt}$, one may extract the spacetime profile of the gauge theory
energy momentum tensor $T_{\mu\nu}$ \cite{Skenderis}
\eq
g_{\mu\nu}(x^\rho,z) =\eta_{\mu\nu}+g^{(4)}_{\mu\nu}(x^\rho)\, z^4+ \OO{z^6}
\quad\quad
\cor{T_{\mu\nu}(x^\rho)}= \f{N_c^2}{2\pi^2}\, g^{(4)}_{\mu\nu}(x^\rho)
\eqx

Let us comment on the possible initial conditions and their interpretation.
On the gravity side, the initial conditions are encoded in a choice of some specific
geometry on a spacelike slice with $\tau=0$. Clearly we have an infinite dimensional space
of possible initial conditions, even if we fix the initial energy density to
a given value. This huge freedom of posing initial conditions
is in marked contrast to hydrodynamics, where just the energy density\footnote{Recall that
we are discussing the boost invariant context.} at some given initial time would suffice.

On the gauge theory side, however, such freedom is also very natural.
If we were at weak coupling, we could imagine preparing the initial state
with various possible momentum distributions of gluons, each initial state
having the same energy density. Then we could let the system evolve and
study the approach of the momentum distributions to a gaussian thermal profile.
The strong coupling analog of this picture would be to start the numerical
evolution from various initial geometries and look for a transition
to an eventual hydrodynamic description. This has been done in \cite{HJW1,HJW2}
starting from $\tau=0$ and using various quite general initial conditions\footnote{Previous
numerical work on boost-invariant plasma started from $\tau=\tau_0>0$
and did not, by design, satisfy energy-momentum conservation in the 
very initial stage \cite{CYBOOST}.}

\section{Results}

The results presented in refs. \cite{HJW1,HJW2} corresponded to initial conditions
with a common fixed energy density at $\tau=0$. Although from the point of view 
of an initial value problem this choice was very natural, it is quite 
inconvenient from the point of view of realistic relativistic heavy-ion
collisions, where we cannot really measure, even indirectly, the energy density
at $\tau=0$. What is known is the temperature at late times (just before freezout) and final
entropy/multiplicity. From hydrodynamic fits one then has a fairly good idea
of the temperature up to the hydrodynamization time, but one cannot
then proceed back to $\tau=0$ without some theoretical assumptions.
Indeed, the AdS results reviewed below indicate that this cannot be
done in a universal way.

\begin{figure}[t]
\centerline{\includegraphics[height=7cm]{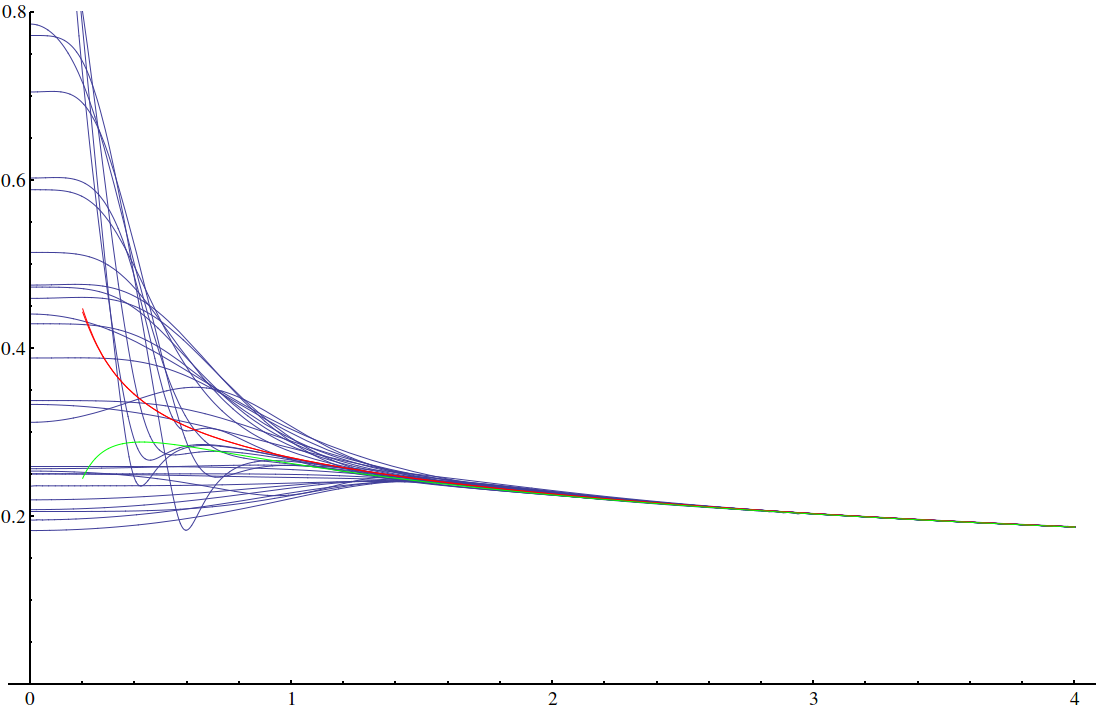}}
\caption{The effective temperature $\teff$ as a function of $\tau$ for the various
initial conditions considered in \cite{HJW1,HJW2}. The green line represents
$3^{rd}$ order hydrodynamic $T_{hydro}^{3^{rd}\, order}(\tau)$, while the red line
corresponds to Borel resummed hydrodynamics $T_{hydro}^{resummed}(\tau)$ (colour online).}
\label{figteff}
\end{figure}

For the above reasons it is interesting to recast the results of the numerical evolution
keeping the large $\tau$ asymptotics (as in (\ref{e.Tnorm})) to be the same for all initial conditions.
The plot of $\teff(\tau)$ for the 29 initial conditions considered is shown in Fig.~\ref{figteff}.

We see exactly the counterpart of the physical picture outlined in section~\ref{s.hydro}. 
At late times all profiles follow very closely hydrodynamic evolution, however
at small times we have a plethora of different curves corresponding
to turning on various nonhydrodynamic degrees of freedom. Moreover, we see that knowing the temperature
at the transition to hydrodynamics gives us no information about the temperature
at $\tau=0$ unless we have some additional information about the initial state.

In order to discuss the finer details of the transition to hydrodynamics,
it is convenient to introduce the dimensionless product $w=\teff \cdot \tau$.
By construction it is independent of any overall scale as well as of the
precise number of degrees of freedom in $\nn=4$ SYM (since we are using
the effective temperature in its definition).

One of the main findings of refs. \cite{HJW1,HJW2} was that even with a very stringent
criterion of transition to hydrodynamics\footnote{I.e. that equations of motion
of $3^{rd}$ order hydrodynamics are satisfied up to $0.5\%$ accuracy.},
a hydrodynamic description holds for $w \geq 0.7$.
Moreover for moderate and large initial data, $w$ at the transition to
hydrodynamics is approximately constant\footnote{Here moderate and large initial
data are distinguished by the value of their initial entropy (see below).}. 
It is interesting to compare this with sample initial data $T_0=500\, MeV, \; \tau_0=0.25\, f\!m$
used for RHIC in \cite{BCFK}, which corresponds to $w=0.63$.
The fact that these values are quite comparable suggests that even
some \emph{quantitative} intuition can be gained from studying strongly coupled $\nn=4$ SYM.
Indeed, from the point of view of our numerical results, it is definitely \emph{not}
surprising that hydrodynamics should be applicable already 
at $T_0=500\, MeV, \; \tau_0=0.25\, f\!m$.

Another key feature which follows, is that at the transiton to hydrodynamics,
the longitudinal and transverse pressures (diagonal components of $T_{\mu\nu}$)
are quite different:
\eq
\Dl p_L \equiv 1-\f{p_L}{\eps/3} \sim 0.7
\eqx
This means that at the transition to hydrodynamics, the plasma is still quite
far from local thermal equilibrium, but this deviation is completely
explained by dissipative hydrodynamic flow. Thus hydrodynamization
is distinct from thermalization and the `early thermalization puzzle'
appears to be really a misnomer.

Finally, we found that despite the varied dynamics apparent in the evolution
profiles seen in Fig.~\ref{figteff}, such key features of the evolution as
entropy production or the hydrodynamization time are strongly correlated
with just a single numerical characterization of the initial state --
an initial entropy\footnote{Initial entropy is characterized geometrically
through an area element of an apparent horizon.}.

Let me now describe some work in progress aiming at the understanding of
the dynamics of the \emph{nonhydrodynamic degrees of freedom} from a 
4D perspective

\section{Beyond hydrodynamics -- towards a four dimensional perspective}

As is clearly seen in Fig.~\ref{figteff}, as well as in the very first
numerical simulations in the boost-invariant context \cite{CYBOOST}, the deviations
from hydrodynamics occur through the appearance of new degrees of freedom in
the plasma system. The appearance of a new degree of freedom is linked to 
the possibility of imposing independent initial conditions for this excitation
(here we think directly in terms of the gauge theory in 4D).
As can be intuitively seen in Fig.~\ref{figteff}, more and more new effective
degrees of freedom appear which reflects the infinite dimensional
freedom of setting up initial conditions on the gravity side.

This infinite set of additional degrees of freedom can be, again intuitively,
understood as being encoded in the additional dimension in the gravitational
description. On the linearized level these degrees of freedom correspond
to black hole quasi-normal modes which have a hierarchy of damping frequencies.
Hence one may expect that in some intermediate region only a couple of
these modes would be relevant. Preliminary analysis of our numerical evolution
profiles compared to resummed hydrodynamics seem to support this conclusion.

A natural question which comes to mind is whether one can describe, perhaps
approximately, the dynamics of such nonhydrodynamic degrees of freedom
purely from a 4D perspective, in a way which agrees with what we know about the
exact numerical description on the AdS side.

Let us recall that hydrodynamics amounts just to the assumption that the energy-momentum
tensor $T_{\mu\nu}$ is completely expressible in terms of a scalar (temperature)
and a unit vector (flow velocity) and an expansion in the number of derivatives. The description of additional nonhydrodynamic degrees
of freedom would entail introducing additional independent effective fields
into  $T_{\mu\nu}$ and providing additional equations of motion.
We hope to report on this issue in a future publication.

\section{Conclusions}

The AdS/CFT correspondence provides a very general framework for studying time-dependent dynamical
evolution of strongly coupled plasma. Using these methods it is possible to follow
the dynamics of the plasma system right from some far from equilibrium configuration
through an initial out-of-equilibrium evolution (or even a collision process 
\cite{CYSW, MHSSW}) onto a very precise
hydrodynamic description. These methods may thus give us qualitative
insight into this very elusive nonperturbative real-time out-of-equilibrium
regime of plasma dynamics. In some cases, once one factors out trivial
differences due to the different number of degrees of freedom, one may
even hope to get some semi-quantitative estimates. However, the key value
of the AdS/CFT methods is that they provide for us a theoretical
laboratory for testing various hypothesis and explicitly showing what kind of
behaviour \emph{could} be possible in a strongly coupled gauge theory --- 
a key example of such an application is the observation that thermalization
is distinct from hydrodynamization, namely 
that the onset of viscous hydrodynamics generically occurs when
the pressures are still quite anisotropic.

\acknowledgments 
\noindent{}This work was supported by NCN grant 2012/06/A/ST2/00396. I would like
to thank the Kobayashi-Maskawa Institute for warm hospitality during the Symposium.

\end{document}